# On the Atomic Structure of Two-Dimensional Materials with Janus Structures

Danil W. Boukhvalov

*College of Science, Institute of Materials Physics and Chemistry, Nanjing Forestry University, Nanjing 210037, P. R. China*
*Theoretical Physics and Applied Mathematics Department, Ural Federal University, Mira Street 19, 620002 Ekaterinburg, Russia*

*The discrepancy between the bright theoretical projections for two-dimensional (2D) Janus structures and the lack of experimental realisation of these structures motivated us to study the effect of structural disorder on the stability of MoSSe, SnSSe, PtSSe, $In_2SSe$ and $GaInSe_2$. The calculation results demonstrate that the difference between metal–sulfur and metal–selenium bonds makes Janus structures frustrated and less energetically favourable than less ordered allotropes of the same compounds. This result explains the difficulties encountered in experimental fabrication of these materials. In the bulk, there is an additional contribution to the total energy from dipole–dipole interactions between layers with a Janus structure that can overcome the energetic cost of structural frustration in layers for compounds with sufficiently large dipole moments. However, the entropic contribution to the free energy decreases the favourability of the ordered Janus structure. The calculation results are used to make recommendations to enable the discovery and synthesis of 2D materials with Janus structures.*

E-mail: danil@njfu.edu.cn

Two-dimensional materials have received tremendous attention from various scientific communities over the last decade. Significant progress has been made in manufacturing 2D materials with various structures [1-3]. Numerous theoretical papers predict various attractive properties for multiple hypothetical 2D structures [1], such as graphene allotropes [4], silicene [5], 2D metal oxides [6], transitional metal dichalcogenides [7] and oxychalcogenides [8]. One of the most spectacular structures that was originally predicted theoretically [9] and realised experimentally four years later [10] was the so-called Janus material, a 2D material with out-of-plane symmetry. The most studied material of this class has been MoSSe (see Fig. 1a and e). Over the last four years, dozens of theoretical studies [6-26] have described the electronic structure and optical, electrical and catalytic properties of various Janus materials. Various dichalcogenides MoXY [11-18], WXY [19-22], SnXY [23-25], PtXY [26,27] and HfXY [28,29] (where X,Y=S, Se and Te) have been common theoretical subjects. Materials with an InSe-like structure (see Fig. 2a and c) have also been considered [31,32].

In the latter class of compounds, both chalcogen atoms in outer layers and atoms in internal layers can be substituted (see Fig. 2b). Despite the bright projections of theoretical papers (these theoretical results are summarized in the mini-review in Ref. [33]) and the growing experience of the scientific community with the synthesis and large-scale production of novel 2D materials [1-8], only two new experimental reports of Janus materials [34,35] have appeared four years after the first synthesis of MoSSe. The growing discrepancy between the rapidly expanding list of theoretical papers on Janus structures and a sole experimental report over the past three years [35] indicates that an important consideration related to the structural stability of Janus 2D materials has been overlooked in all the aforementioned theoretical papers.

In the theoretical studies cited above, the stability of Janus compounds was checked by calculating phonon spectra [11,15,21,25,27,28,32], as well as by first principles molecular dynamics calculations in some studies [21,28]. Vacancy formation has also been simulated in several studies. It would appear that all the steps of the standard procedure for assessing the stability of an atomic structure have been performed. However, the lack of experimental confirmation of theoretical predictions is a tip-off that an important step has not been performed in all these theoretical studies. To identify the missing element in the theoretical modelling of 2D structures, we first analyse the experimental approach to synthesizing Janus materials. In contrast to widely used methods for making 2D materials (delamination from the bulk or vapour deposition on a surface), MoSSe monolayers have been obtained by etching the upper layer of sulfur atoms from a deposited $MoS_2$ monolayer and then attaching a Se layer over the MoS substrate [10]. This state-of-the-art approach is far from the standard procedure used to fabricate 2D materials, in which the obtained atomic structure is determined only by the thermodynamics and kinetics of the process. In the aforementioned approach, only the substrate (the MoS layer) determines the formation of the final atomic structure. Thus, the possibility of the formation of different atomic structures with the same crystal structure and chemical composition should be checked. In this study, we investigate the effect of disorder in the distribution of selenium and sulfur atoms in MoSSe and similar compounds with and without chalcogen vacancies on the total energy of these systems. Compounds with InSe-like structures and differences between monolayers and bulk for the investigated structures were also taken into consideration and analysed.

To maintain realistic computational costs for modelling disordered structures, we chose 4×4×1 $MoS_2$-like (Fig. 1) and InSe-like (Fig. 2) supercells. Theoretical modelling of GaN layers on various substrates was performed using the SIESTA pseudopotential code [36], where the generalised gradient approximation (GGA-PBE) [37] was employed for the exchange-correlation potential in a spin-polarised mode and correction of van der Waals forces [38]. Full optimisation of the atomic positions was carried out, during which the electronic ground state was consistently found using norm-conserving pseudopotentials [39] for the cores and a double-ξ-plus polarisation basis of localised orbitals. The forces and total energies were optimised with accuracies of 0.04 eV Å$^{-1}$ and 1.0 meV/cell (or less than 0.02 meV/atom), respectively.

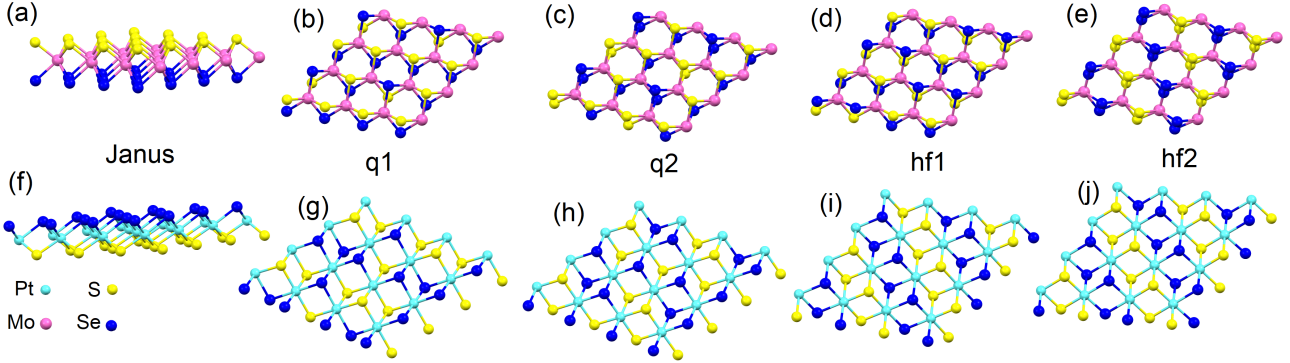

**Figure 1.** Optimised atomic structures of various structural configurations of *2H*-MoSSe (a–d) and *1T*-PtSSe (f–j).

**Table 1.** Calculated differences between the lowest enthalpy and enthalpy of various structural configurations (see Fig. 1) for monolayers with and without anionic defects (the bulk data are shown in parentheses). The ground state is considered to have zero enthalpy. The rightmost column shows the temperatures at which the contributions to the Gibbs free energy from the configurational entropy and the difference between the enthalpies of the ground-state and Janus structures are equal in magnitude.

| Lattice Type | Compound | Janus | q1 | q2 | hf1 | hf2 | T, K |
|---|---|---|---|---|---|---|---|
| 2H | MoSSe<br>MoS$_{0.94}$Se$_{0.94}$ | 35.91 (0.00)<br>33.54 (0.00) | 9.46<br>7.92 | 12.25<br>9.96 | 0.00 (3.52)<br>0.00 (0.67) | 5.59<br>33.94 | 73 (—)<br>68 (—) |
| 1T | MoSSe<br>MoS$_{0.94}$Se$_{0.94}$ | 455.81<br>431.96 | 442.72<br>437.36 | 475.58<br>464.09 | 354.88<br>354.89 | 498.32<br>479.94 | 930<br>882 |
| 1T | SnSSe<br>SnS$_{0.94}$Se$_{0.94}$ | 26.38 (17.80)<br>16.20 (10.39) | 9.01<br>3.49 | 9.09<br>6.14 | 0.00 (0.00)<br>0.48 | 0.00<br>0.00 (0.00) | 54 (36)<br>33 (21) |
| 1T | PtSSe<br>PtS$_{0.94}$Se$_{0.94}$ | 26.07<br>18.53 | 25.41<br>18.64 | 4.92<br>0.24 | 0.00<br>0.00 | 0.00<br>0.69 | 53<br>38 |
| InSe | In$_2$SSe<br>In$_2$S$_{0.94}$Se$_{0.94}$ | 19.68<br>20.19 | 4.69<br>4.70 | 7.39<br>7.84 | 0.00<br>0.00 | 3.36<br>6.74 | 40<br>41 |
| InSe | GaInSe$_2$<br>GaInSe$_{1.88}$ | 47.66<br>0.00 | 17.94<br>46.65 | 30.35<br>53.52 | 0.00<br>58.92 | 3.37<br>52.44 | 97<br>— |

The first step in our modelling procedure was to compare the difference in the total energies of the Janus structure of *2H*-MoSSe with those of different less ordered structures. For this purpose, we interchanged the positions of a quarter and half of the S and Se atoms in different ways (see Fig. 1b-e). We denote these structures by q1 and q2 (for replacement of a quarter of the atoms) and hf1 and hf2 (for replacement of half the atoms). Note that the number of Mo, S and Se atoms in all the structures remained the same. For all the considered structures, we fully optimised the atomic positions and lattice parameters. The calculation results (see Table 1) demonstrate that the lowest total enthalpy corresponds to the hf1 structure, followed closely by the hf2 structure, both of which are more energetically favourable than the Janus structure. The enthalpies of the q1 and q2 structures lie between the two aforementioned extremes. Note that the hf1 and hf2 structures are not a real ground state for MoSSe but only the closest possible approximation to the ground state for a supercell of this size. However, the closeness of the total enthalpies of the hf1 and hf2 structures suggests that a further increase in the disorder in the distribution of S and Se atoms will lead to an insignificant decrease in the total enthalpy; therefore, the structures with the lowest total enthalpy among those considered are considered as the ground state. To eliminate concerns about the effect of the supercell size on the results, we performed calculations for several quasi-disordered structures for a MoSSe 6×6×1 supercell of 108 atoms. The calculated total energies per formula unit for these disordered configurations were similar to those for hf1 and hf2 presented in Table 1 and 32.4 and 30.8 meV smaller than that of the Janus structure. The reported numbers are very close to those obtained for smaller supercells (see Table 1). Hence, a 4×4×1 supercell can be considered a suitable model system.

The next step in our modelling procedure was to expand the list of considered compounds and structures. For this purpose, we performed similar simulations for the *1T*-MoSSe structure, as well for the *2H* and *1T* structures of SnSSe (see Fig. 1e-j) and PtSSe. The calculation results (see Table 1) demonstrate that, as for $MoS_2$ and $MoSe_2$, the *2H* configuration is significantly more favourable than the *1T* configuration for MoSSe. Note that the Janus structure is also the least energetically favourable of all five considered configurations for *1T*-MoSSe. In the case of *1T*-SnSSe and *1T*-PtSSe, in the absence of Se and S vacancies, the hf2 structures have the same total enthalpies as the hf1 structures, which are also considerably smaller than those of the Janus structures. In the case of the *2H* configurations of the considered compounds, the Janus structure is stable in the absence of vacancies or other structural disorder. The presence of these imperfections leads to a spontaneous transition from the *2H* to the *1T* structure. This instability results from the very large energy difference between the *2H* and *1T* configurations (approximately 0.8 eV/unit for SnSSe and approximately 1.2 eV/unit for PtSSe vs. 0.4 eV/unit for MoSSe).

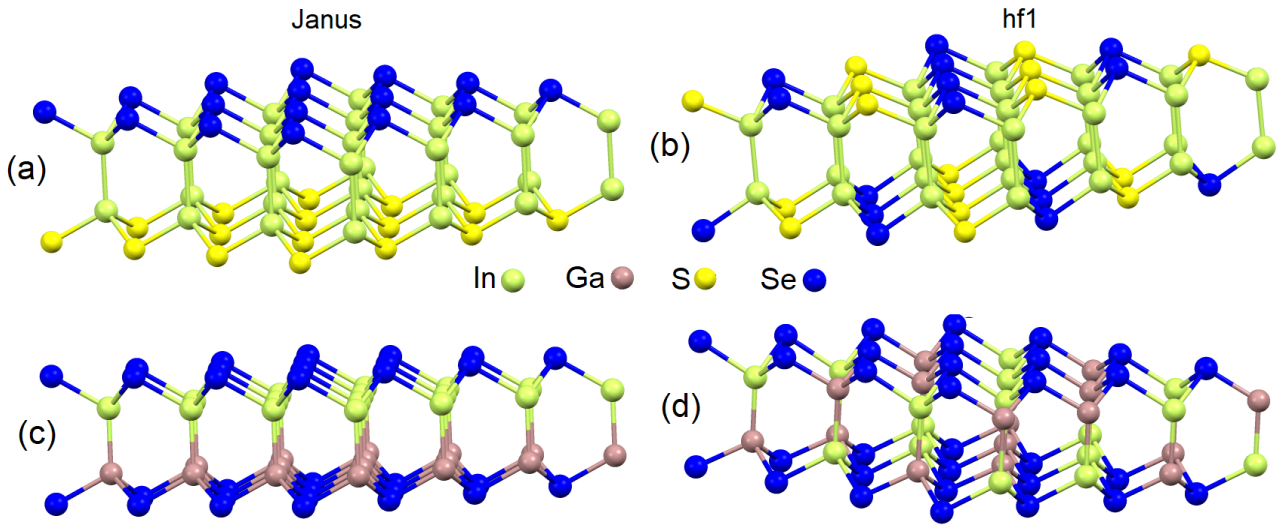

**Figure 2.** Optimised atomic structure of Janus (a and c) and ground-state hf1 (b and d) structures for $In_2SSe$ (a and b) and $InGaSe_2$ (c and d).

The next step in our analysis was to investigate the observed unfavourability of the Janus–structure for different types of 2D structures that have been reported in the literature [25,26]. For this purpose, we used InSe-based structures. Note that Janus structures can be built from these types of layers by replacing chalcogenide atoms, as discussed above, or by replacing atoms in the inner layer. The formation of both types of InSe-based Janus structures was considered. For this purpose, $In_2SSe$ and $InGaSe_2$ monolayers were built (see Fig. 2a and b). Disordered structures similar to the aforementioned disordered structures were considered for both compounds. As for the dichalcogenides, a disordered hf1 structure was found to possess the lowest energy.

As the formation of chalcogen vacancies is highly probable, the role of these defects was also studied. For this purpose, one sulfur atom and one selenium atom (or two selenium atoms for $InGaSe_2$) located in a similar position near the same metal centre were removed from all considered supercells, followed by optimisation of the atomic structure and lattice parameters of the supercells. The calculation results (see Table 1) demonstrate that the presence of chalcogen vacancies can stabilise the Janus structure in only one compound ($InGaSe_{1.88}$), where the Janus structure corresponds to the inner part of the membrane. In other structures, the presence of vacancies of the S-Se pair does not change the preferability of disordered structures, such as hf1 and hf2, over ordered Janus structures.

**Table 2.** Calculated in-plane lattice vector and dipole moment for Janus and ground-state structures (hf1 or hf2, see Table 1) for monolayers of all considered compounds with and without defects.

| Compound | In-plane Lattice Vector, Å | | Dipole Moment, Debyes per Cell | |
|---|---|---|---|---|
| | Janus | Ground State | Janus | Ground State |
| MoSSe | 3.272 | 3.262 | 3.372 | 0.003 |
| $MoS_{0.94}Se_{0.94}$ | 3.227 | 3.221 | 3.170 | 0.217 |
| SnSSe | 3.790 | 3.786 | 1.601 | 0.017 |
| $SnS_{0.94}Se_{0.94}$ | 3.816 | 3.810 | 1.645 | 0.037 |
| PtSSe | 3.769 | 3.761 | 4.369 | 0.003 |
| $PtS_{0.94}Se_{0.94}$ | 3.694 | 3.698 | 3.874 | 0.200 |
| $In_2SSe$ | 3.946 | 3.960 | 2.157 | 0,008 |
| $In_2S_{0.94}Se_{0.94}$ | 3.927 | 3.927 | 2.072 | 0.089 |
| $GaInSe_2$ | 3.926 | 3.920 | 1.309 | 0.038 |
| $GaInSe_{1.88}$ | 3.872 | 3.872 | 0.911 | 0.911 |

The calculation results presented above lead to the conclusion that the formation of the Janus structure in dichalcogenides is energetically unfavourable. This unfavourability is attributed to the mismatch between the length of the metal–S and metal–Se bonds. For example, the Mo–S bond in $MoS_2$ (2.417 Å) is visibly shorter than the Mo–Se bond in $MoSe_2$ (2.547 Å). The formation of the Janus structure creates structural frustration induced by different metal-chalcogen-metal bonds on different sides of the membrane. Alternating between Janus and hf1 or hf2 structures increases the structural disorder, which in turn leads to a decrease in the structural frustration and, therefore, the total enthalpy of the system. The formation of nanotubes, such as MoSSe single wall nanotubes [40], is another way to reduce structural frustration in 2D Janus structures.

**Table 3.** The change in the total energy in meV/bond of *1T*-MoSSe/ *1T*-$MoS_{0.94}Se_{0.94}$ after an out-of-plane shift of 0.04 Å of one Se or S atom in the supercell (dE(Se) and dE(S), respectively) and after a 2% uniaxial in-plane extension of the supercell (last column).

| Structural Configuartion | dE (Se) | dE (S) | dE (strain) |
|---|---|---|---|
| Janus | 9.0/9.0 | 11.1/10.6 | 11.42/10.21 |
| ground state (see Tab. 1) | 10.7/10.2 | 10.9/10.4 | 10.06/10.28 |

The considered changes in the atomic structures of monolayers do not lead to significant changes in the planar lattice parameters for all the considered compounds. The changes in the lattice parameters caused by the transition between the Janus and disordered structures are smaller than those caused by the formation of chalcogen vacancies (see Tab. 2). Therefore, the main experimental methods used to characterise atomic structures (XRD, XPS, XANES, and HRTEM) cannot distinguish the Janus structure from the disordered structure. To determine the difference in the vibrational properties of the Janus and ground-state structures, we calculated the energies required for an out-of-plane shift of 0.04 Å of sulfur and selenium and in-plane uniaxial extension of a MoSSe supercell with and without defects. The calculation results shown in Table 3 demonstrate that the difference between the energies required for in- and out-of-plane vibrations of the supercell with a Janus structure and in the ground state are of the same order as changes in the same energies caused by the formation of vacancies. Two conclusions can be drawn from these results. First, the vibrational entropic term is almost constant for all the considered structures. Second, vibrational spectroscopy cannot distinguish Janus structures with vacancies from other disordered structures. The Janus and disordered structures also have almost the same electronic structures (see Fig. 3). The key difference between the macroscopic physical properties of Janus and less ordered structures is the zero dipole moment of membranes with structures such as hf1 and hf2.

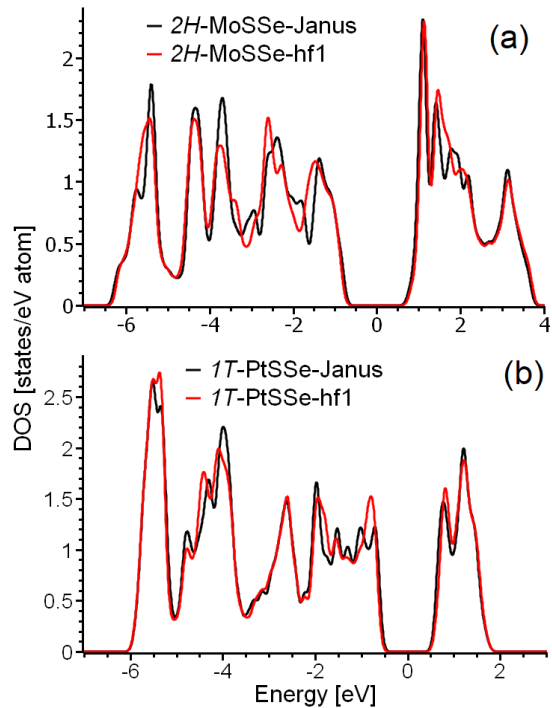

**Figure 3.** Calculated total densities of states for Janus (black curves) and ground-state hf1 structures (red curves) of *2H*-MoSSe (a) and *1T*-PtSSe (b). The Fermi energy is set to zero.

The last step of our analysis was to study how the transition from monolayer to bulk affected the energetics of the considered structures. For this purpose, bulk structures of AB-stacked *2H*-MoSSe and AA-stacked *1T*-SnSSe were simulated. Janus and hf1 configurations were considered for both structures. The calculation results demonstrate that the Janus structure is 3.52 meV/unit more favourable than the hf2 structure for bulk *2H*-MoSSe. The hf1 structure remains favourable for bulk *1T*-SnSSe, but the magnitude of the difference in energies between the hf1 and Janus structures is less than that for the monolayer (17.80 meV/unit vs. 26.38 meV/unit). To understand these results, we also calculated the binding energies for all four considered structures. The binding energies are 259 and 221 meV/unit for the Janus and hf1 structures of *2H*-MoSSe2, respectively. The binding energies of the Janus and hf1 structures for *1T*-SnSSe2 are 362 and 354 meV/unit, respectively. Based on these numbers, we propose that the favourability of configurations in the bulk is determined by interplay between the energy cost for the formation of a monolayer and the binding energies between the layers of the considered type. In the case of *2H*-MoSSe, the energy cost for forming the Janus structure (35.91 meV/unit) overlaps with the energy gain from the change in the binding energy (38 meV/unit). Note that the difference between these two energies is almost the same as the difference between the enthalpies of the Janus and hf1 structures (3.52 meV/unit). The difference between the enthalpies of the Janus and hf1 structures for *1T*-SnSSe is 26.38 meV/unit. The difference between the binding energy for these two structures in the bulk phase is only 8 meV/unit. This value is also almost the same as the difference between the enthalpies of the hf1 and Janus structures for the bulk and monolayer (8.58 meV/unit). To explain the observed difference between *2H*-MoSSe and *1T*-SnSSe, it is necessary to identify the nature of the chemical bonds between the layers. First, we note that the magnitudes of binding energies in both cases are on the order of hundreds of meVs, which is an order of magnitude larger than the energies of so-called van der Waals bonds. Thus, we propose that the bonds in dichalcogenides are more electrostatic than van der Waals bonds. The key difference between Janus and hf1 or hf2 structures is the zeroing of the dipole moment in disordered structures (see Table 2) for equal numbers of sulfur and selenium atoms on both sides of the membrane. Note that the magnitude of the dipole moment for the Janus structure of *2H*-MoSSe is almost twice that for *1T*-SnSSe. This difference in the dipole moments corresponds to the difference in the binding energies between the hf1 and Janus structures for *2H*-MoSSe and *1T*-SnSSe (38 and 8 meV/unit, respectively). These observations lead to the conclusion that the layers in the bulk phase are bound by electrostatic and van der Waals forces for the considered structures. The electrostatic interactions are higher in magnitude than the van der Waals forces. The intrinsic electrical polarisation of the layers leads to an increase in the electrostatic interactions between the layers. To verify this assumption, we performed similar calculations for the compound with the dipole moment with the largest magnitude in the monolayer with the Janus structure (*1T*-PtSSe, see Table 2). The calculation results demonstrate that the bulk phase with the Janus structure is significantly energetically favoured (12.16 meV/unit) over other disordered structures. These robust electrostatic interactions make mechanical or microwave-assisted exfoliation of materials with Janus structures difficult. However, liquid-phase exfoliation in nonpolar solvents could be successful.

Only changes in the enthalpies of the systems were considered in the aforementioned analyses. However, the transition from an ordered Janus structure to disordered hf1 or hf2 structures corresponds to an increase in configurational entropy, as described by the formula

$$S = k_B \ln W,$$

where W is the number of possible configurations. The value of W can be calculated using combinatorics,

$$W = n!/(n-2)! = (n-1)n,$$

where n is the number of units in the supercell. For the 4×4×1 supercells of 18 formula units considered in this study, the decrease in the configurational entropy as a result of a transition from hf1 or hf2 to Janus structures is approximately 0.5 meV/K per formula unit. Thus, even at room temperature, the entropic contribution (approximately 148 meV/unit) exceeds the energetic gain from hf1-to-Janus structural transitions, even in PtSSe. A larger 4×4×1 supercell of 36 formula units was used to determine the effect of the size of the supercell, and the configurational entropic contribution at room temperature was slightly higher than 184 meV/unit. This result explains why it is difficult to form Janus structures and the fabrication of these structures is rarely successful. As the contribution from vibrational entropy changes is negligible (see the discussion presented above), only the configurational entropy contributes noticeably to the free energies of the compounds. The calculated configurational entropy can be used to evaluate the temperature at which the dH and TdS terms in the Gibbs free energy are equal in magnitude:

$$dG = dH + TdS,$$

where dH is the difference between the enthalpy of the Janus and ground-state configurations. The calculated temperatures (see the rightmost column in Table 1) are below 100 K for all considered realistic configurations and show that the entropic contribution can be decreased by decreasing the synthesis temperature.

The calculated results show that for all the considered structures in monolayers, Janus structures are less energetically favourable than disordered structures. The energetic disadvantage of Janus structures results from the high energetic cost of structural frustration caused by noticeable differences in the metal–S and metal–Se bond lengths on different sides of the Janus layers. The entropic contribution makes the formation of ordered Janus structures less energetically favourable than ground-state structures. However, the formation of disordered structures results in the decrease and zeroing of dipole moments in the considered monolayers. The calculated results also demonstrate that for the bulk phase of the considered materials, the interplay between the energetic cost of structural frustration within layers and the energetic gain from additional dipole–dipole interactions and the entropic contribution determine the most energetically favourable structure.

The simulation results are used to make the following recommendations for fabricating 2D materials with Janus structures: (i) compounds with large dipole moments should be chosen, (ii) a bulk phase should be grown and subjected to exfoliation, (iii) the use of an external electric field or polar substrate is strongly recommended, (iv) the synthesis temperature should be as low as possible to reduce the entropic contribution, and (v) syntheses should be performed under chalcogen-rich conditions because the presence of cationic vacancies reduces the dipole moment (see Table 2). The approach used to model Janus structures and their disordered allotropes can be used for prediction and subsequent machine-learning-based screening of the most promising materials with Janus structures for synthesis and large-scale production.

**Acknowledgements**
The author acknowledged support Ministry of Science and Higher Education of the Russian Federation (through the basic part of the government mandate, Project No. FEUZ-2020-0060) and Jiangsu Innovative and Entrepreneurial Talents Project.

**References**

1) N. Briggs, S. Subramanian, Z. Lin et. al., *2D Mater.*, 2019, **6**, 022001.
2) A. Qadir, T.K. Le, M. Malik, K.A.A. Min-Dianeye, I. Saeed, Y. Yu, J.R. Choi, P.V. Pham, *RSC Adv.*, 2021, **11**, 23860-23880.
3) H. Zhang, M. Chhowalla, Z. Liu, *Chem. Soc. Rev.*, 2018, **47**, 3015-3017.
4) Y.X. Yu, *J. Mater. Chem. A*, 2013, **1**, 13559–13566.
5) Y.X. Yu, *J. Phys. Chem. C*, 2019, **123**, 205-213.
6) J.H. Li, J. Wu, Y.X. Yu, *J. Phys. Chem. C*, 2020, **124**, 9089-9098.
5) H. Zhang, M. Chhowalla and Z. Liu, Chem. Soc. Rev., 2018, **47**, 3015-3017.
7) Y.X. Yu, Appl. Surf. Sci. 2021, **546**, 149062.
9) Y. C. Cheng, Z. Y. Zhu, M. Tahir, U. Schwingenschlögl, *EPL* , 2003, **102**, 57001.
10) A.-Y. Lu, H. Zhu, J. Xiao, et. al., *Nat. Nanotech.* 2017, **12**, 744-751.

11) [MoSSe, WSeTe, E, M, phon] V.V. Thanh, N.D. Van, D.V. Truong, R. Saito, N.T. Hung, *Appl. Surf. Sci.* 2020, **526**, 146730.
12) C. Jin, X. Tang, X. Tan, S.C. Smith, Y. Dai, L. Kou, *J. Materi. Chem. A*, 2019, **7**, 1099-1106.
13) A.C. Riis-Jensen, M. Pandey, K.S. Thygesen, *J. Phys. Chem. C*, 2018, **122**, 24520-24526.
14) J. Wang, H. Shu, T. Zhao, P. Liang, N. Wang, D. Cao, X. Chen, *Phys.Chem.Chem.Phys.*, 2018, **20**, 18571.
15) M. Yagmurcukardes, C. Sevik, F.M. Peeters, *Phys. Rev. B*, 2019, **100**, 045415.
16) Y. Ji, M. Yang, H. Lin, T. Hou, L. Wang, Y. Li, S.-T. Lee, *J. Phys. Chem. C*, 2018, **122**, 3123-3129.
17) W.-J. Yin, B. Wen, G.-Z.g Nie, X.-L. Wei,L.-M. Liu, *J. Mater. Chem. C*, 2018, **6**, 1693.
18) Z. Cui, N. Lyu, Y. Ding, K Bai, *Physica E*, 2021, **127**, 114503.
19) W. Zhou, J. Chen, Z. Yang, J. Liu, F. Ouyang, *Phys. Rev. B*, 2019, **99**, 075160.
20) D. Er, H. Ye, N.C. Frey, H. Kumar, J. Lou, V.B. Shenoy, *Nano Lett.* 2018, **18**, 3943-3949.
21) A. Kandemir and H. Sahin, *Phys.Chem.Chem.Phys.*, 2018, **20**, 17380.
22) Y. Chen, J. Liu, J. Yu, Y. Guo, Q. Sun, *Phys.Chem.Chem.Phys.*, 2019, **21**, 1207.
23) S.-D. Guo, X.-S. Guo, R.-Y. Han, Y. Deng, *Phys.Chem.Chem.Phys.*, 2019, **21**, 24620.
24) P. Wang, Y. Zong, H. Liu, H. Wen, H.-X. Deng, Zh. Wei, H.-B. Wu, J.-B. Xia, *J. Phys. Chem. C*, 2020, **124**, 23832-23838.
25) H.T.T. Nguyen, V.V. Tuan, C.V. Nguyen, H.V. Phuc, H.D. Tong, S.-T. Nguyen, N.N. Hieu, *Phys.Chem.Chem.Phys.*, 2020, **22**, 11637.
26) R. Peng, Y. Ma, B. Huang, Y. Dai, *J. Mater. Chem. A*, 2019, **7**, 603.
27) D.D. Vo, T.V. Vu, S. Al-Qaisi, H.D. Tong, T.S. Le, C.V. Nguyen, H.V. Phuc, H.L. Luong, HR. Jappor, M.M. Obeid, N.N. Hieu, *Superlat. Microstr.*, 2020, **147**, 106683.
28) D.M. Hoat, M. Naseri, N.N. Hieu, R. Ponce-Perez, J.F. Rivas-Silvag, Tuan V. Vu, G.H. Cocoletzi, *J. Phys. Chem. Solids*, 2020, **144**, 109490.
29) M. Barhoumi, K. Lazaar, S. Bouzidi, M. Said, *J. Molec. Graph. Model.*, 2020, **96**, 107511.
30) T.V. Vu, H.D. Tong, D.P Tran, N.T.T. Binh, C.V. Nguyen, H.V. Phuc, H.M. Do, N.N. Hieu, *RSC Adv.*, 2019, **9**, 41058.
31) T.V. Vu, V.T.T. Vi, C.V. Nguyen, H.V. Phuc, N N Hieu, *J. Phys. D: Appl. Phys.* 2020, **53**, 455302.
32) A. Kandemir and H. Sahin, *Phys. Rev. B*, 2018, **97**, 155410.
33) R. Li, Y. Cheng, W. Huang, *Small,* 2018, **14**, 1802091.
34) J. Zhang, S. Jia, I. Kholmanov, L. Dong, D. Er, W. Chen, H. Guo, Z. Jin, V.B. Shenoy, L. Shi, J. Lou, *ACS Nano*, 2017, **11**, 8192-8198.
35) H. Li, Y. Qin, B. Ko, D.B. Trivedi, D. Hajra, M.Y. Sayyad, L. Liu, S.-H. Shim, H. Zhuang, S. Tongay, *Adv. Mater.*, 2020, **32**, 2002401.
36) J.M. Soler, E. Artacho, J.D. Gale, A. Garsia, J. Junquera, P. Orejon, D. Sanchez-Portal, *J. Phys.: Condens. Matter.*, 2002, **14**, 2745.
37) J.P. Perdew, K. Burke, M. Ernzerhof, *Phys. Rev. Lett.*, 1996, **77**, 3865.
38) H. Rydberg, M. Dion, N. Jacobson, E. Schröder, P. Hyldgaard, S. I. Simak, D. C. Langreth, B.I. Lundqvist, *Phys. Rev. Lett.*, 2003, **91**, 126402.
39) O.N. Troullier, J.L. Martins, *Phys. Rev. B*, 1991, **43**, 1993.
40) A.E.G. Mikkelsen, F.T. Boolle, K.S. Thygesen, T. Vegge, I.E. Castelli, *Phys. Rev. Mater.*, 2021, **5**, 014002.